\documentclass[aps,showpacs,amsmath,amssymb,preprint,prb,floatfix,superscriptaddress,a4]{revtex4}

\usepackage[dvips]{graphicx}
\usepackage{amsmath}
\newcommand{\vecto}[1]{\mathrm{\textbf{#1}}}  

\begin{document}

\title{Effects of screened Coulomb impurities on autoionizing two-electron resonances in spherical quantum dots}

\author{Michael Genkin}
\affiliation{Atomic Physics, Stockholm University, AlbaNova, S-10691 Stockholm, Sweden}

\author{Eva Lindroth}
\affiliation{Atomic Physics, Stockholm University, AlbaNova, S-10691 Stockholm, Sweden}

\date{\today}

\begin{abstract}
In a recent paper (Phys. Rev. B {\bf 78}, 075316 (2008)), Sajeev and Moiseyev demonstrated that the bound-to-resonant transitions and lifetimes of 
autoionizing states in spherical 
quantum dots can be controlled by varying the confinment strength. In the present paper, we report that 
such control can in some cases be compromised by the presence of Coulomb impurities.
It is demonstrated that a screened Coulomb impurity placed in the vicinity of the dot center can lead to bound-to-resonant transitions and to avoided 
crossings-like behavior when the screening of the impurity charge is varied.
It is argued that these properties also can have impact on electron transport through quantum dot arrays. 

\end{abstract}

\pacs{32.80.Zb, 73.21.La}

\maketitle

\section{Introduction}
Autoionizing states have been very thoroughly investigated in conventional atomic systems. The accuracy achieved nowadays in experiment and theory is 
as high as to resolve hyperfine splittings in dielectronic resonances~\cite{Lest_08}. Another, currently very active, research topic involving 
autoionizing states are pump-and-probe experiments with short laser pulses, which make it possible to follow the autoionization process in real time 
and thus to resolve electron dynamics in atoms~\cite{Uibe_07,Zhel_09}.
However, the autoionization process in artificial atoms like quantum dots is less known;
From an experimental point of view, it is much more difficult to observe autoionization, since, contrary to "usual" atomic systems, 
quantum dots are imbedded into the semiconductor. The conventional atomic approach where autoionizing states are revealed as resonances in the cross 
section spectra for ionization by photon or particle impact is, in the case of quantum dots, not easily employed. 
Also theoretical data is rather scarce, since  
the confinment of the electrons in the dot is often modelled by a harmonic potential, which 
automatically excludes the possibility of resonances, while in reality the confining potential is, of course, finite. 
To circumvent this problem, other confinment models were suggested to study the autoionization process in two-electron quantum dots, like a finite 
well~\cite{Bucz_96,Byli_05,Ferr_09} or a Gaussian potential~\cite{Adam_00,Saje_08}, which indeed led to interesting observations such as 
resonance-induced enhancement of the dot sensitivity to photons~\cite{Saje_08} or entanglement in resonances~\cite{Ferr_09}.
In the present work, we aim to study the effects of screened Coulomb impurities on the positions and lifetimes of such autoionizing resonances. The 
impact of charged impurities on the properties of quantum dots was addressed in connection with different confinment models, like
a parabolic potential~\cite{Mukh_97,Lee_00,Yau_03,Aich_05} or an infinite well~\cite{Chuu_92} but also finite 
potentials~\cite{Pand_04,Sahi_05,Kass_07,Sahi_08,Sahi_08_E}. 
Another interesting aspect is the behavior of quantum dots with impurities in external fields, which can give rise to effects like emergence of 
stable non-dispersive electron wave 
packets~\cite{Kali_05}, field-enhanced electron localization~\cite{Lee_98} or change of optical properties~\cite{Bask_07}.
However, we are only aware of one paper explicitly investigating the role of impurities in the autoionization process~\cite{Bucz_96}. 
Therein, Buczko and Bassani considered a finite potential well with a hydrogenic impurity and chose an analytical method based on scattering theory 
techniques. Here, we adopt a different approach commonly known as "complex scaling" or "complex coordinate rotation" to determine the positions and 
widths of the resonances in a Gaussian-shaped spherical two-electron quantum dot with a Coulomb impurity. 
The method is described in more detail in the following section~\ref{mod} along with the computational procedure, and the results are presented in 
section~\ref{results}. A discussion and conclusions are given in section~\ref{concl}.

\section{Method}\label{mod}
The Hamiltonian of a spherical Gaussian two-electron quantum dot with a Coulomb impurity reads
\begin{equation}
H=-\sum_{i=1}^{2}\left(\frac{\hbar^2\nabla^2_{q_i}}{2m^{*}}+U_0e^{-\alpha q_i^2}+\frac{\eta e^2}{4\pi\epsilon_0\epsilon q_i}\right)
+\frac{e^2}{4\pi\epsilon_0\epsilon\left|\vecto{q}_1-\vecto{q}_2 \right|},
\end{equation}
where $\vecto{q}_1,\vecto{q}_2$ are the coordinates of the two electrons, $m^{*}$ the effective electron mass,
$\epsilon$ the dielectric constant of the 
semiconductor, $U_0$ the depth of the confining potential, $\alpha$ a parameter describing the range of the latter and $\eta$ the effective charge of 
the impurity. For convenience, we introduce scaled parameters as suggested in Ref.~\cite{Saje_08}:
\begin{eqnarray}\nonumber
\vecto{r}_i&=&\frac{m^*}{m_e\epsilon}\vecto{q}_i,\\
\label{scale}
V_0&=&\frac{m_e\epsilon^2}{m^*}U_0,\\
\nonumber
\beta&=&\frac{m_e^2\epsilon^2}{\left(m^*\right)^2}\alpha,
\end{eqnarray}
where $m_e$ is the electron mass, so that the Hamiltonian can be written as:
\begin{equation}\label{Ham}
H=\frac{m^*}{m_e\epsilon^2}
\left[-\sum_{i=1}^{2}\left(\frac{\hbar^2\nabla^2_{r_i}}{2m_e}+V_0e^{-\beta r_i^2}+\frac{\eta e^2}{4\pi\epsilon_0 r_i}\right)
+\frac{e^2}{4\pi\epsilon_0\left|\vecto{r}_1-\vecto{r}_2\right|}\right].
\end{equation}
From here on, all quantities will be given in effective atomic units, and the values for 
the corresponding semiconductor can be reobtained from Eqs.~(\ref{scale}). For example, in the case of GaAs we have $\epsilon=12.4$ and 
$m^*/m_e=0.067$, which gives the effective energy unit $1\,{\rm Ha}^*\approx 11.857\,{\rm meV}$ and the effective length 
unit $a_0^*\approx 9.794\,{\rm nm}$.
To determine the bound as well as the resonant states of the system, we diagonalize the complex scaled Hamiltonian within a B-Spline 
basis set. This method was earlier applied to describe autoionizing resonances in atomic~\cite{Bran_99,Bran_02} and 
exotic~\cite{Lind_03,Lind_04_E,Genk_09} systems.
The Hamiltonian~(Eq.(\ref{Ham})) is dilation-analytic, so that the uniform complex scaling~\cite{Agui_71,Bals_71,Simo_72,Mois_98} of the coordinates 
can be imposed:
\begin{equation}
\vecto{r}_i\rightarrow\vecto{r}_i e^{i\theta},
\end{equation}
with a real parameter $0<\theta<\pi/4$.
The eigenvalues of the scaled Hamiltonian are complex. However, the physical states are invariant with respect to $\theta$ while pseudo-continuum 
states are rotated into the complex plane by the angle $2\theta$. After diagonalizing the scaled Hamiltonian for different values of $\theta$, one 
obtains the bound states of the system as the eigenvalues with vanishing imaginary parts and the resonant states as complex eigenvalues,
\begin{equation}
E_{\rm res}=E_{\rm pos}-i\frac{\Gamma}{2},
\end{equation}
the real part of which gives the position of the resonance and the negative imaginary part the halfwidth of the latter. It is connected to the lifetime 
$\tau$ as $\tau=\hbar/\Gamma$. 
The diagonalization of the scaled Hamiltonian is carried out numerically using LAPACK-routines, and below we briefly describe the computational 
procedure. 
The radial part of the one-particle Hamiltonian $h$ is given by
\begin{equation}\label{honepart}
h=\frac{m^*}{m_e\epsilon^2}\left(-\frac{\hbar^2}{2m_e}\frac{\partial^2}{\partial r_i^2}+\frac{l_i(l_i+1)\hbar^2}{2m_e r_i^2}-
V_0e^{-\beta r_i^2}-\frac{\eta e^2}{4\pi\epsilon_0r_i}\right),
\end{equation}
and in the first step, the corresponding radial Schr{\"o}dinger equation 
is solved in a sufficiently large box using piecewise polynomial functions $B_k$ (so called B-Splines~\cite{Boor_78,Bach_01} of a certain order $k$) 
defined on a given knot sequence. Thus, the radial one-particle eigenfunctions $\varphi_j$ are obtained in the form
\begin{equation}
\varphi_j(r)=\sum_{k}c_{kj}B_k(r)
\end{equation}
with expansion coefficients $c_{kj}$. Subsequently, the matrix elements of the interaction potential 
\begin{equation}
\frac{m_e\epsilon^2}{m^*}V_{12}=\frac{e^2}{4\pi\epsilon_0\left|\vecto{r}_1-\vecto{r}_2\right|}
\end{equation}
between the obtained states
are computed using multipole expansion, where the radial integration can be carried out to machine accuracy using Gaussian quadrature,  
while the angular integration is performed analytically using Racah algebra~\cite{Lind_86}.
The full Hamiltonian is then set up in the basis of coupled eigenstates to the one-particle Hamiltonians under consideration of the 
Pauli principle. Complex scaling is imposed 
by multiplying the kinetic terms by $\exp(-2i\theta)$ and the Coulomb-terms by $\exp(-i\theta)$. The real and imaginary part of the complex rotated
Gaussian confining potential are given by
\begin{eqnarray}
{\rm Re}\left[-V_0\exp\left(-\beta \left(r_ie^{i\theta}\right)^2\right)\right]&=&-V_0\exp\left(-\beta r_i^2\cos(2\theta)\right) 
\cos\left(\beta r_i^2\sin(2\theta)\right)\\
{\rm Im}\left[-V_0\exp\left(-\beta \left(r_ie^{i\theta}\right)^2\right)\right]&=&V_0\exp\left(-\beta r_i^2\cos(2\theta)\right)
\sin\left(\beta r_i^2\sin(2\theta)\right)
\end{eqnarray}
and are scaled accordingly. After the setup of the Hamiltonian matrix is completed, the latter is diagonalized in the final step.

\section{Results}\label{results}
In the present calculations, we considered both donor and acceptor impurities, and therefore we varied the effective charge in the regions 
$\eta>0$ and $\eta<0$. This is a quite simple picture of the screening mechanism, but it should be sufficient to illustrate the effect of impurities on 
autoionizing states. In a more advanced approach, one could also include a possible spatial dependence of the screening. For example, 
Kwon~\cite{Kwon_06} recently presented a model where the screening is modelled by an exponentially 
decreasing potential
\begin{equation}
V_{\rm imp}^{i}=\frac{q}{4\pi\epsilon\epsilon_0r_i}\exp\left(-\frac{r_i}{r_s}\right),
\end{equation}
where $q$ is the true impurity charge and $r_s$ the screening length which depends on the doping concentration in the semiconductor and the 
temperature (see Eqs.(9)-(15) in Ref.~\cite{Kwon_06}). 
We simplify the treatment by 
varying $\eta$ independently of $r_i$, as it is often done in atomic many-electron systems to model the screening effects by core electrons. 
Also, since we are mostly interested in autoionizing states which are situated not very close to the dot center, this approximation seems reasonable.

\begin{figure}
\scalebox{0.55}{\includegraphics{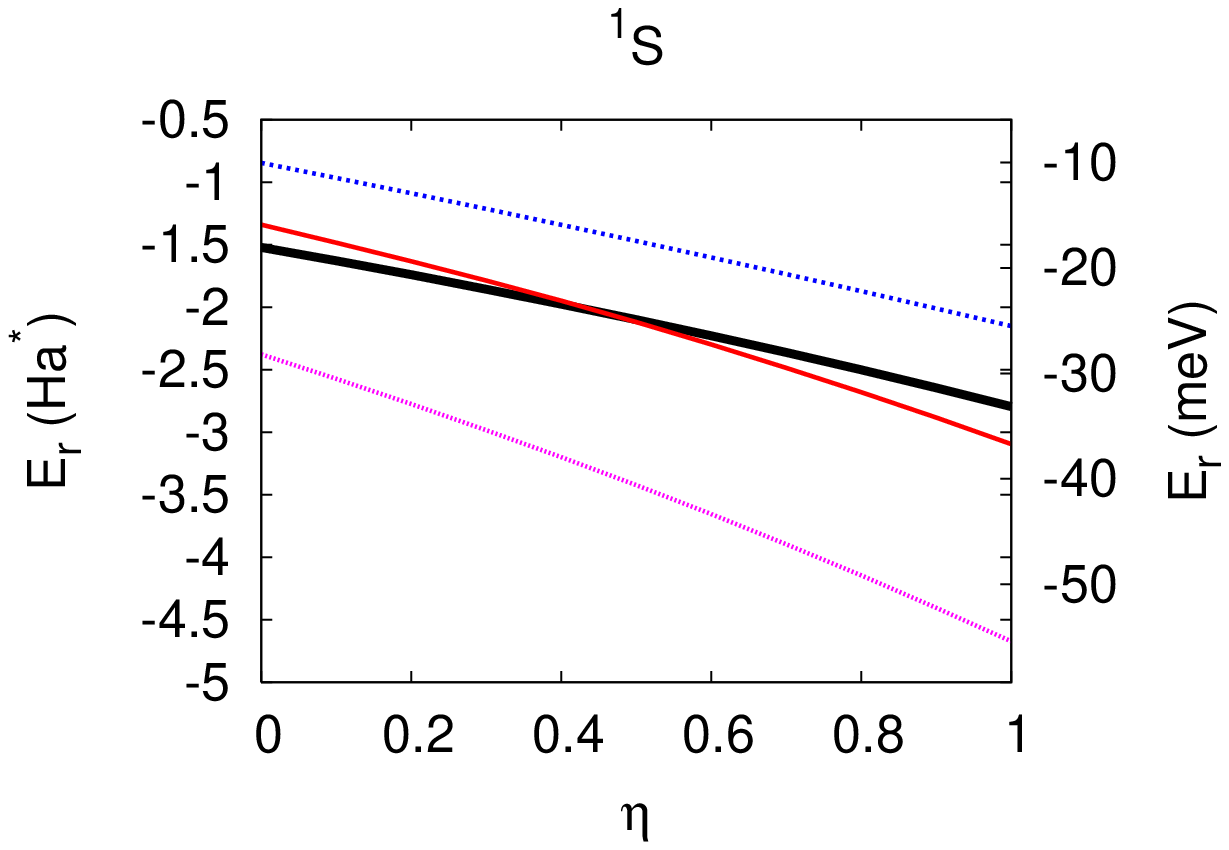}}
\scalebox{0.55}{\includegraphics{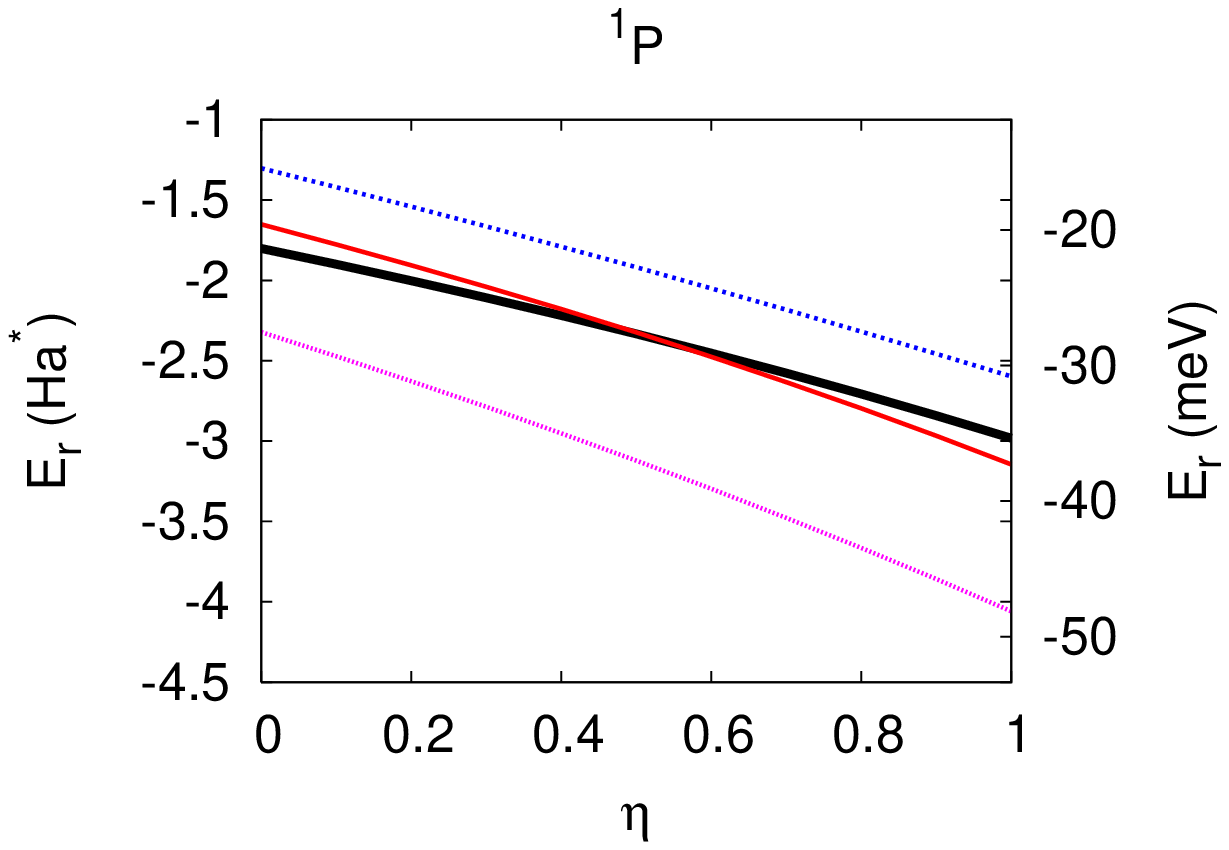}}\\
\scalebox{0.55}{\includegraphics{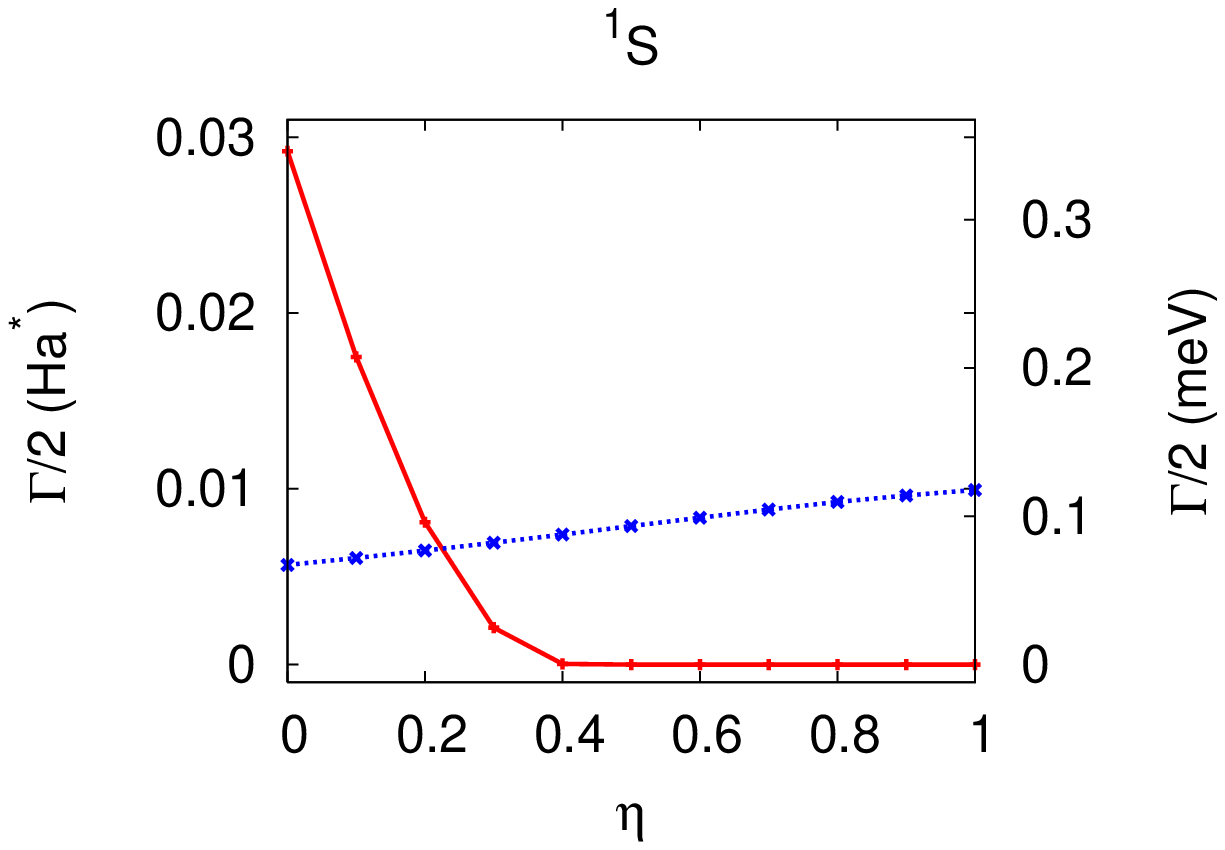}}
\scalebox{0.55}{\includegraphics{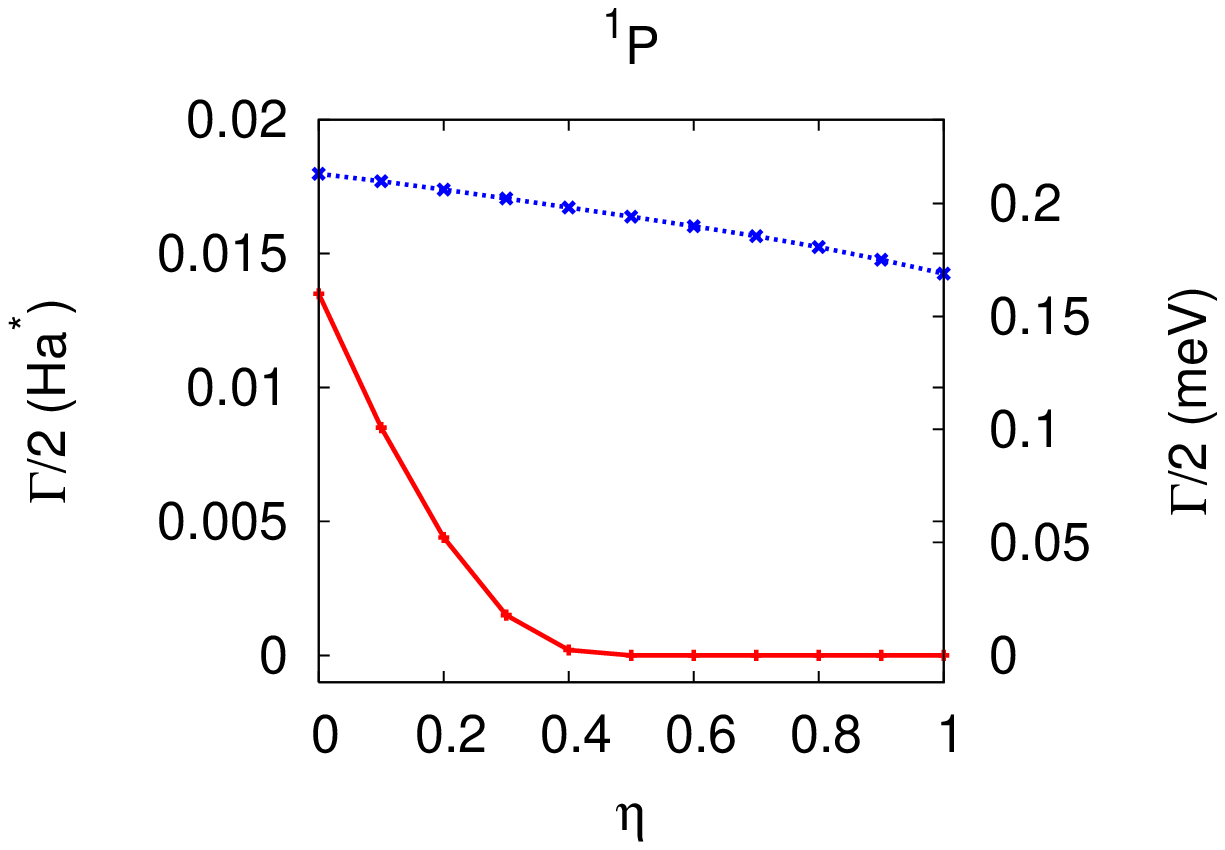}}
 \caption{(Color online) Donor impurity ($\eta>0$). Upper panels: energy positions of the three lowest states (colored lines) and the position of the 
$1s$-threshold (solid black 
line) as the function of the screening strength, shown for $^1S$ (left) and $^1P$ (right) symmetry. Lower panels: Halfwidths of the occuring resonant 
states. As the position of the second state (red line) crosses the threshold, it becomes a bound state and its width vanishes. 
The values for the range parameter were taken as $\beta=0.21\,(a_0^*)^{-2}$ for the $^1S$ symmetry and $\beta=0.13\,(a_0^*)^{-2}$ for the $^1P$ 
symmetry. The energies and halfwidths are given both in scaled Hartree units (left axis) and meV (right axis) with material parameters of GaAs.
\label{fig1}}
\end{figure}

\begin{figure}
\scalebox{0.55}{\includegraphics{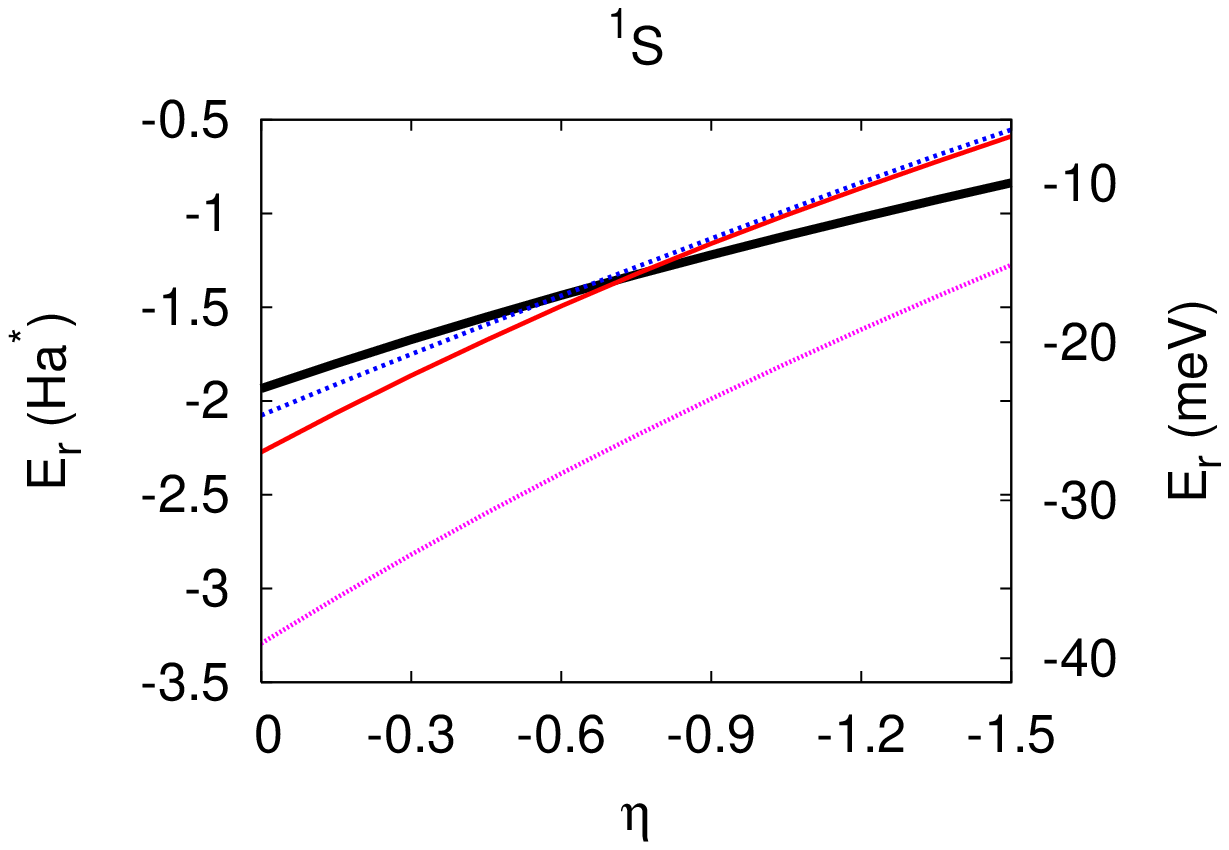}}
\scalebox{0.55}{\includegraphics{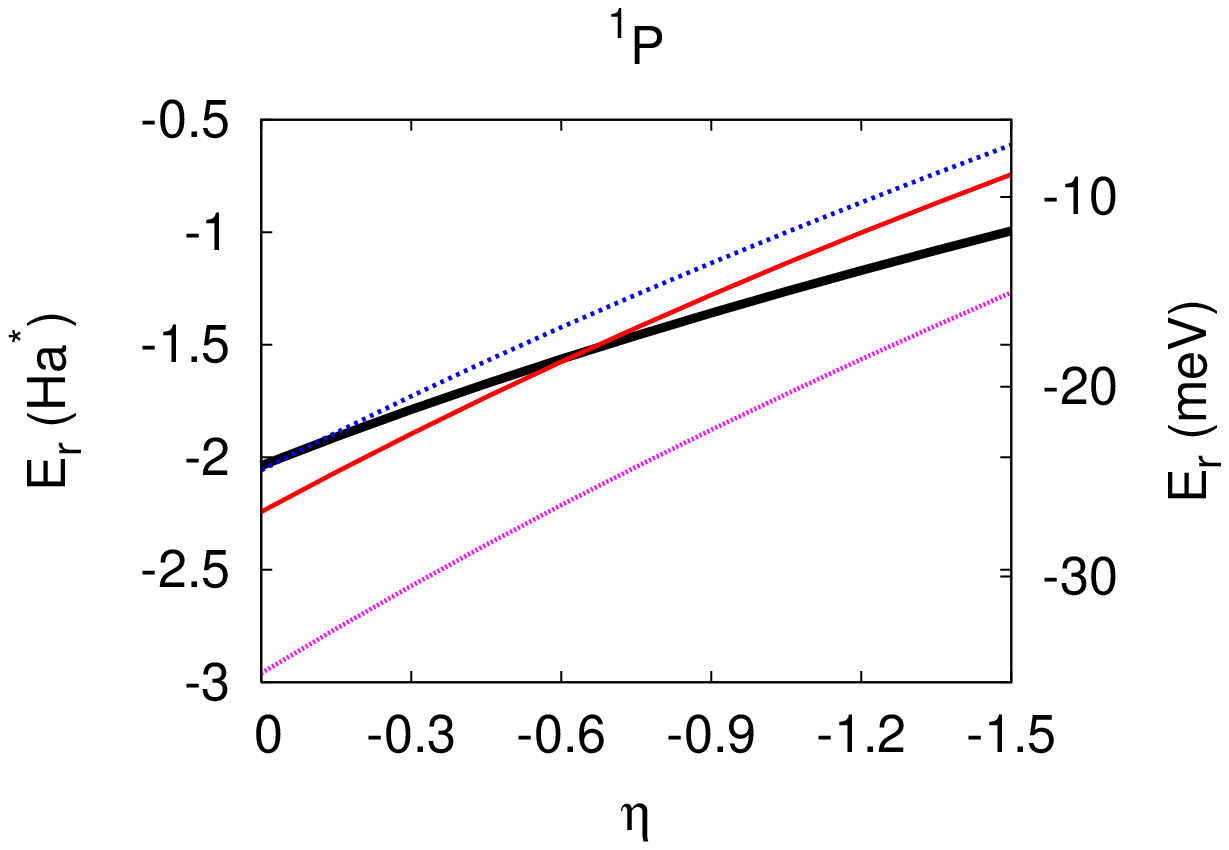}}\\
\scalebox{0.55}{\includegraphics{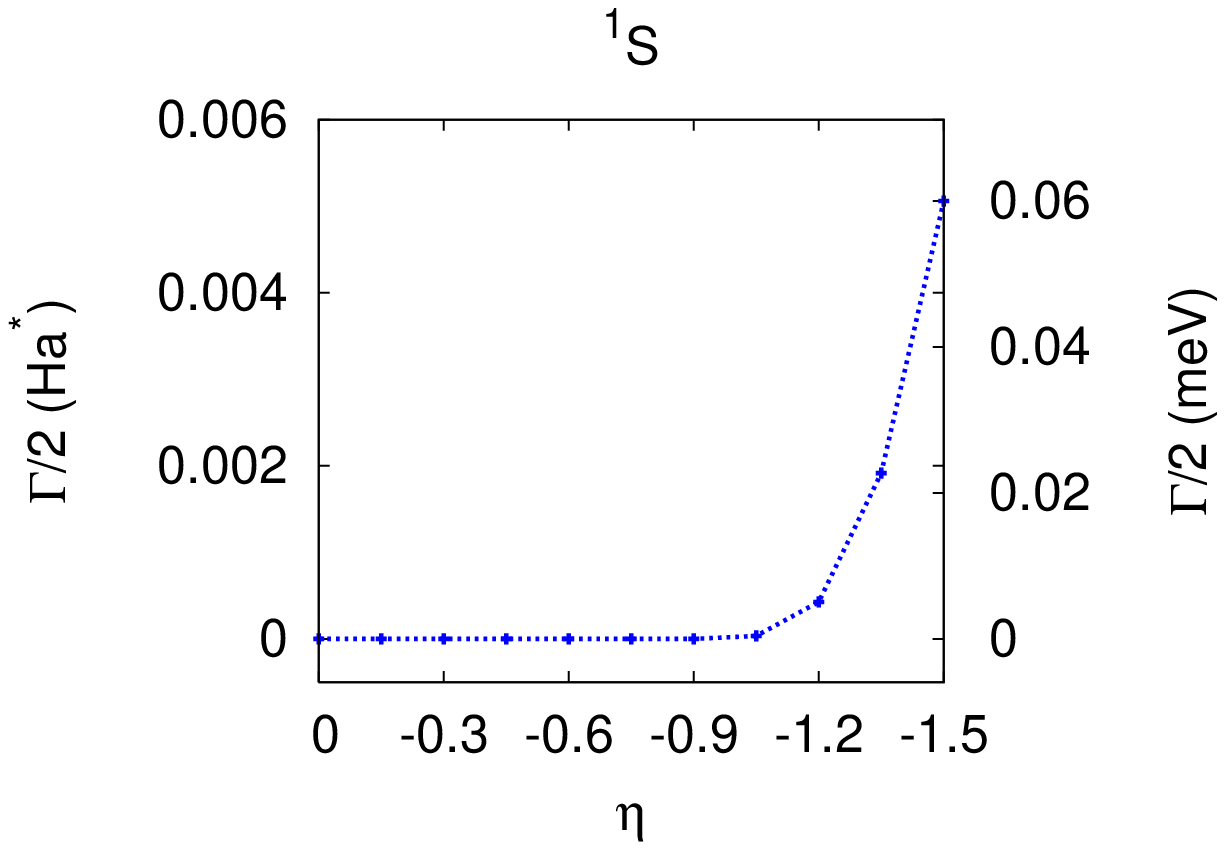}}
\scalebox{0.55}{\includegraphics{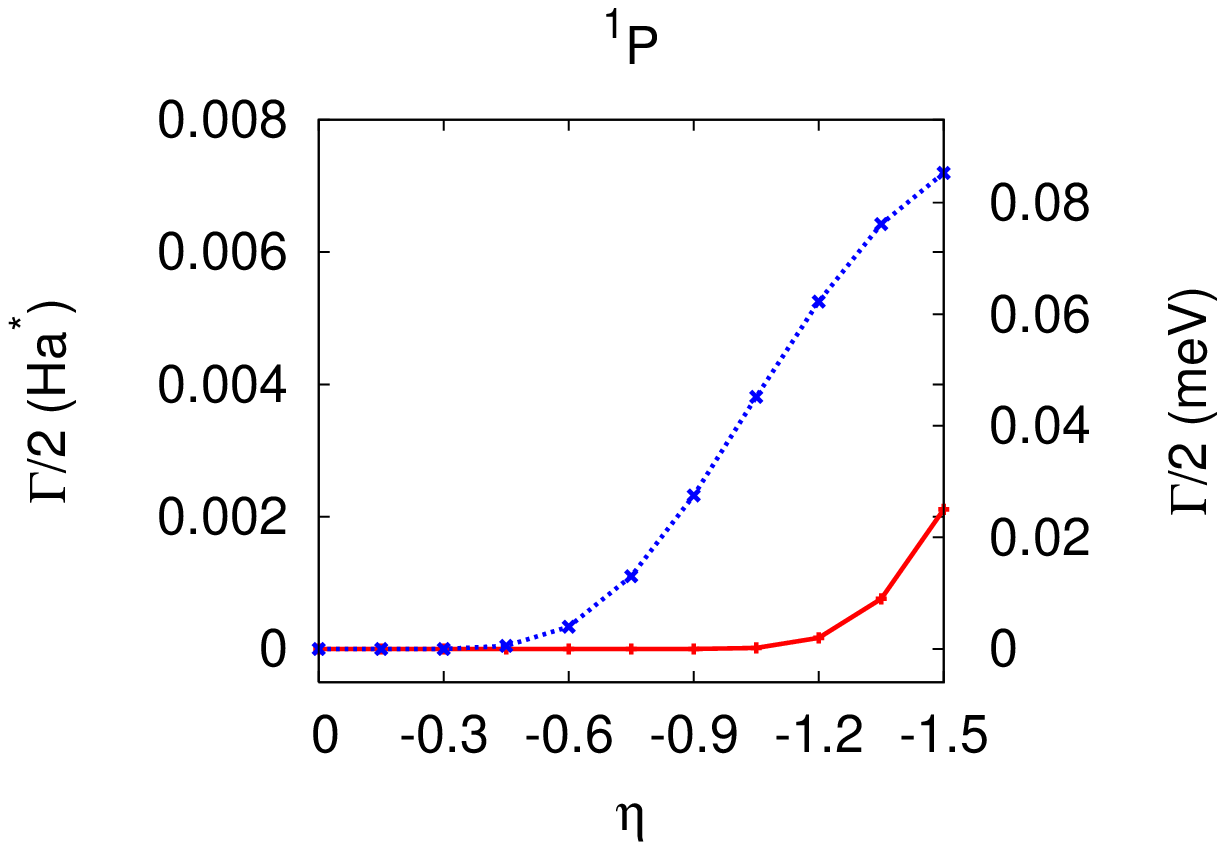}}
 \caption{(Color online) Same as in Fig.~\ref{fig1} for an acceptor impurity ($\eta<0$). Here, we observe that two states 
(blue dotted and red solid lines) cross the threshold. However, in the case of $^1S$ symmetry, the width of the second state (red solid line) is orders 
of magnitude smaller than the one of the third state and is not shown in the plot since it is not distinguishable from zero on the given scale.
The values for the range parameter were taken as $\beta=0.10\,(a_0^*)^{-2}$ for the $^1S$ symmetry 
and $\beta=0.08\,(a_0^*)^{-2}$ for the $^1P$ symmetry. \label{fig2}}
\end{figure}

\begin{figure}
\scalebox{0.55}{\includegraphics{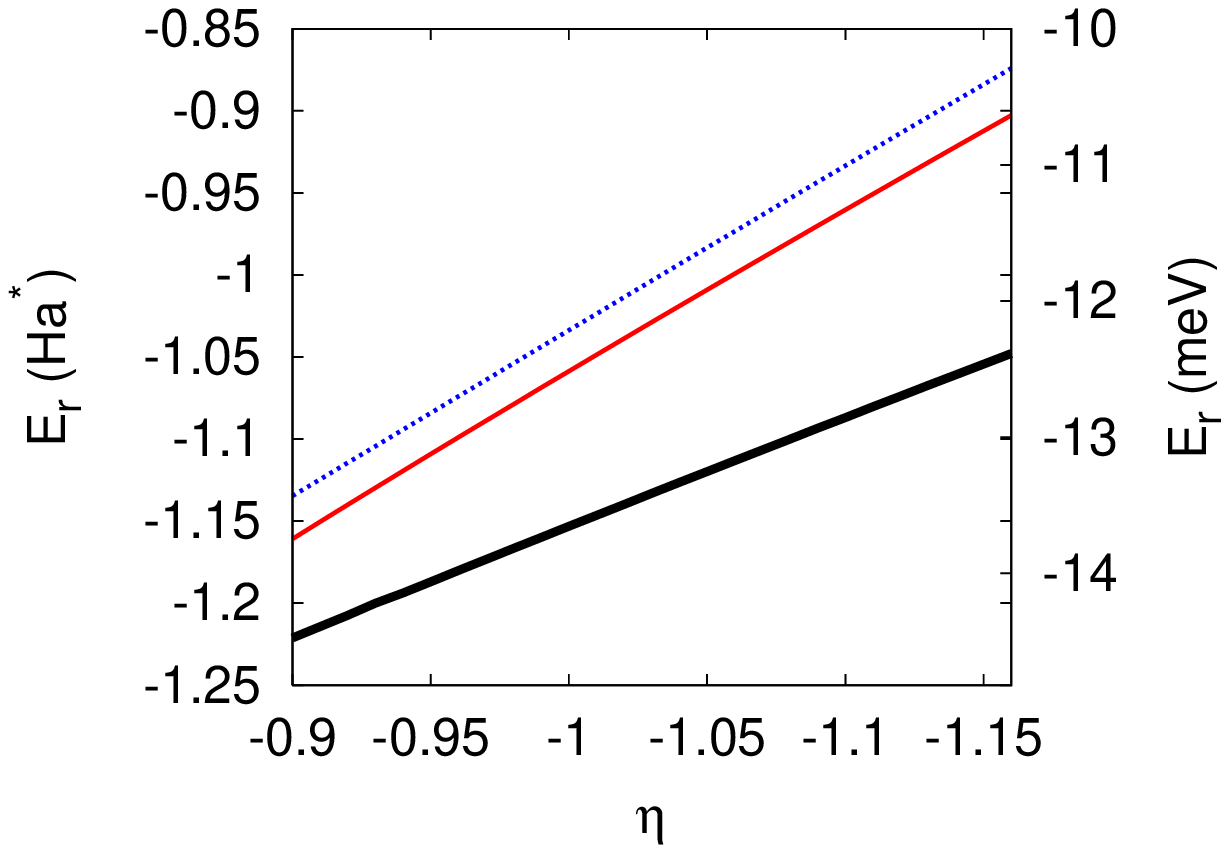}}
\scalebox{0.55}{\includegraphics{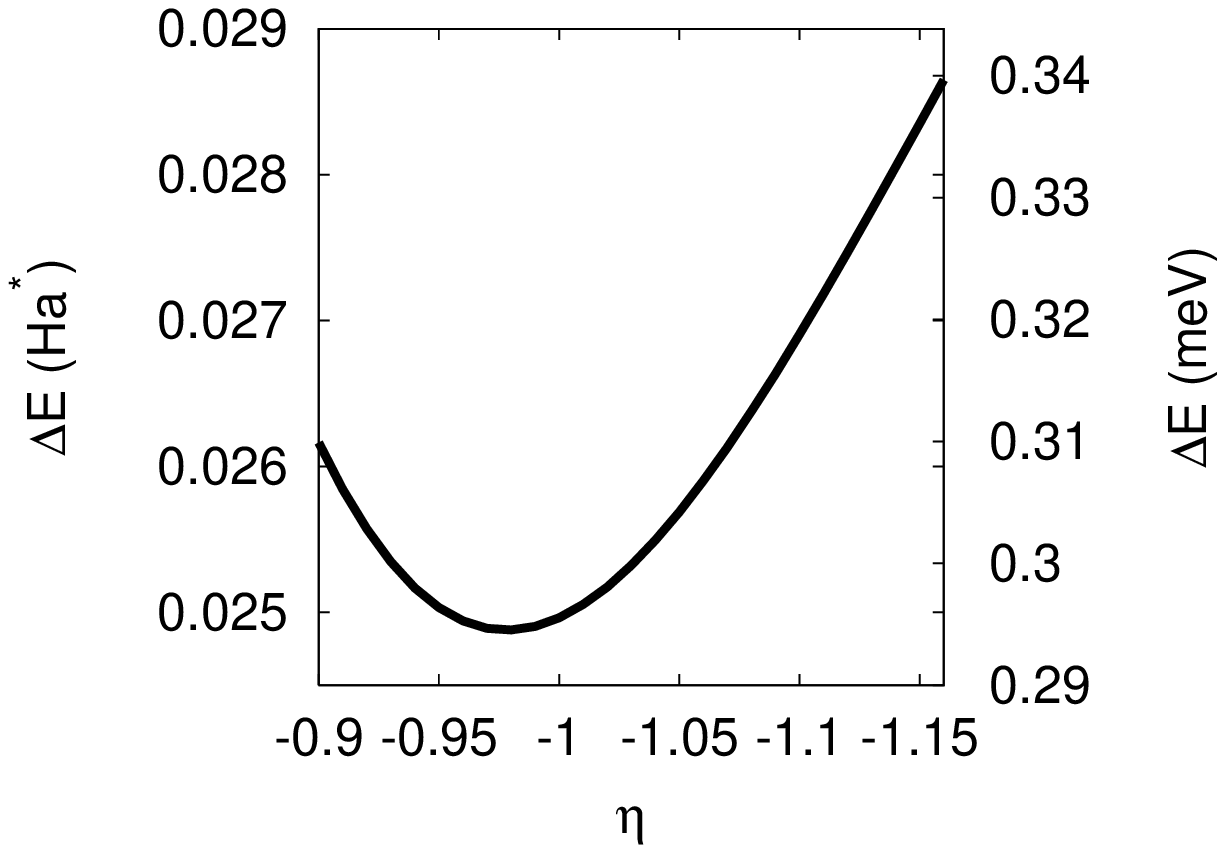}}
 \caption{(Color online) Left panel: "zoom-in" into the plot shown in the upper left panel of Fig.~\ref{fig2}. Right panel: energy difference between 
the two states. The behavior is similiar to an avoided crossing. The range  parameter of the confinment is $\beta=0.10\,(a_0^*)^{-2}$ 
 \label{fig3}}
\end{figure}

The parameters of the confining potential are chosen in the same region as suggested in 
Ref.~\cite{Saje_08}: throughout the calculations, the potential depth is kept fixed at $V_0=3\,{\rm Ha}^*$ and for the range parameter $\beta$ we take 
certain values which are well suited to illustrate the physical behavior we aim to demostrate. To represent the basis states, a sequence of 48 knot 
points with a box size of $R=24\,a_0^*$ is used to generate the B-Spline set, the order of which is $k=7$ throughout the paper. We restrict our 
treatment to singlett resonances and include all configurations of $s-s,p-p$ and $d-d$ type for the $^1S$ symmetry and all configurations 
of $s-p$ and $p-d$ type for the $^1P$ symmetry, which is enough to reach sufficient convergence. 
The numerical stability of the method was confirmed by successfully reproducing the positions and 
widths of the resonances as given in Ref.~\cite{Saje_08} for the case $\eta=0$ obtained with a Gaussian basis set. 
Figures~\ref{fig1} and~\ref{fig2} show the positions and halfwidths of the three lowest states and the $1s$-threshold for donor and acceptor impurities 
with different screening strengths.  
We observe that a donor impurity can turn an autoionizing state into a bound state, while for an acceptor impurity the opposite behavior 
is seen. Somewhat similar situations are known to occur in atomic systems, when a certain electronic configuration gives or does not give rise to 
autoionizing resonances depending on the nuclear charge. For example, the $(1s^22p^2)^1S$ state is autoionizing in beryllium~\cite{Lind_96}, while in 
the case of any other beryllium-like ion (e.g. beryllium-like carbon~\cite{Mann_98}) this state is bound.
Of course, this analogy does not fully hold in the case studied here, since the confining potential remains unchanged, 
but to some extent it allows a qualitative insight in the observed behavior.
Another interesting aspect in the case of donor impurities is that the width of the state which remains autoionizing (blue curve in Fig.~\ref{fig1}) 
is affected differently 
for different total angular momenta: in the case of $^1S$ symmetry it is slightly increasing while for the $^1P$ symmetry it is slightly decreasing as 
the effective impurity charge grows. Furthermore, we would like to point out that, for an acceptor impurity, the positions of the second and third 
states become very close to each other after crossing the threshold (blue and red curves in the upper left panel in Fig.~\ref{fig2}). By 
"zooming in" into the relevant region, we see that these curves show an avoided-crossing like behavior. It is illustrated in Fig.~\ref{fig3},
where we also plot the energy difference $\Delta E$ of the states vs. the effective impurity charge.
To summarize, we observe that the dot spectra are quite sensitive with respect to impurities, both concerning the positions and, in case of 
autoionizing states, also the lifetimes, in particular because impurities can cause threshold-crossings so that bound states become resonant or vice 
versa. In the following section~\ref{concl}, we discuss the physical implications of the observed behavior which allows us to draw some conclusions.

\section{Discussion and conclusions}\label{concl}
In the context of the presented results, we focus on two main topics in our discussion: resonance-enhanced sensitivity of the quantum dots to 
photons and the role of autoionizing resonances in transport processes through quantum dot chains. The practical purpose behind controlling the 
positions and lifetimes of resonances in quantum dots by adjusting the confining potential~\cite{Saje_08} was a possible 
application of the latter as sensitive photodetectors. As demonstrated therein, the presence of an autoionizing resonance leads, in fact, to a very 
significant increase of the photoionization rate, since autoionizing states can be intermediately populated in the photoionization process. Our 
results, however, indicate that in case of such an application attention should be paid to the purity of the semiconductor, since a donor impurity 
could turn an autoionizing state into a bound state which would considerably decrease the detector sensitivity. In other words, donor impurities 
counteract the controlled efficiency of such photodetectors. 
As for the role of acceptor impurities, we would like to mention their possible impact on electron transport 
through quantum dot chains. Let us for example imagine an array of quantum dots, prepared in a way that each of them initially contains one electron 
and consider the propagation of an electronic wave packet from one end of the chain to another~\cite{Niko_04}. If, in such a situation, acceptor 
impurities are present in one or several dots and give rise to autoionizing resonances, it could lead to an additional channel for quantum transport 
where the electron is captured into a resonant state and remains there for a time span comparable to the lifetime of the resonance before it is 
released back to the continuum. This mechanism would thus compete with tunneling bewtween coupled quatum dots, possibly even giving rise to 
interference effects among the propagation paths. Qualitatively, one may even compare the situation to the first step in the process of dielectronic 
recombination in ions, when free electrons are captured into doubly excited states by simultaneous
excitation of a core electron. Of course, in a semiconductor the resonant charecterstic would be less pronounced since the propagating electrons are 
not monochromatic; nevertheless, such a parallel between "usual" and artificial atoms is quite intriguing.

In conclusion, we studied autoionizing resonances in the presence of Coulomb impurities in spherical Gaussian-shaped two-electron quantum dots using 
the complex 
scaled direct diagonalization method. We found that donor impurities can turn resonant states with a finite lifetime into bound states, while acceptor 
impurities have the opposite effect. Implications of these features were discussed in the context of photoionization and transport processes in quantum 
dots, underlining the importance of the semiconductor purity in these particular applications.

\section*{Acknowledgements}
We thank Prof. Nimrod Moiseyev for stimulating discussions during the 2009 meeting of the COST-action and Dr. Luca Argenti for helpful remarks.
This work was financially supported by the G{\"o}ran Gustafsson Foundation and the Swedish research council (VR).

\end{document}